\documentclass[11pt]{article}
\usepackage[utf8]{inputenc}
\usepackage{amsmath,amssymb,graphicx}
\usepackage{hyperref}
\usepackage[T1]{fontenc}
\usepackage{mathpazo}
\usepackage[font=small,labelfont=bf]{caption}
\usepackage{longtable}
\usepackage{verbatim}
\usepackage{amsfonts}
\usepackage{cite}
\usepackage{float}
\usepackage[utf8]{inputenc}
\usepackage{multirow}
\usepackage{booktabs}
\usepackage{pdflscape}

\newcommand{\beq}{\begin{eqnarray}}
\newcommand{\eeq}{\end{eqnarray}}

\usepackage[table,usenames,dvipsnames]{xcolor}
\usepackage{framed}
\usepackage{afterpage}

\author{Eliott Rosenberg and JiJi Fan \\
{\em Department of Physics, Brown University, Providence, RI 02912}}
\title{\bf Cooling in a Dissipative Dark Sector}
\begin{document}
\maketitle

\begin{abstract}
The possibility of a subdominant component of dark matter dissipating energy could lead to dramatic new phenomenology such as the formation of a dark disk. One rigorous way to assess this possibility and settle the debate on its feasibility is to include the dissipative dark component in a numerical hydrodynamical simulation. A necessary input to such a simulation is a prescription including energy dissipation rates of different processes and rates of processes that change the number densities of dark ions and atoms. In this article, we study the simplest dissipative dark sector which consists of a dark electron and proton, both charged under a dark gauged $U(1)$. We present approximate analytic formulas for energy loss rates due to Compton scattering, bremsstrahlung, recombination, collisional ionization and collisional excitation as well as the rates of number density change. We also include the heating rate due to photoionization. The work serves as the first step to realize a numerical simulation including a dissipative dark sector, which hopefully can shed more light on the formation and properties of a dark disk originating from dark matter self-interactions. 
\end{abstract}

\section{Introduction}
\label{sec:intro}

The existence of dark matter is one of the greatest mysteries in the Universe. Very little about it is known: it is mostly cold, collisionless, and does not interact directly with light. Therefore, an increasing number of theoretical possibilities have been considered. One pressing question facing dark matter hunters is whether there could be new possibilities leading to novel phenomenology and search strategies. Here, we revisit and relax some traditional assumptions, namely that dark matter is single component and thus should have no dissipative dynamics due to the halo shape constraints. Indeed as shown in Ref.~\cite{Fan:2013yva, Fan:2013tia}, if dark matter consists of multiple species, a subdominant component is still allowed to have dissipative dynamics, possibly resulting in a dark disk with interesting consequences. Possible effects of such a dark disk and variants of the scenario have been studied further in Refs.~\cite{CyrRacine:2012fz, McCullough:2013jma, Fan:2013bea, Fischler:2014jda, Randall:2014lxa, Randall:2014kta, Reece:2015lch, Kramer:2016dqu, Kramer:2016dew, Agrawal:2017rvu}. 

One missing piece in the study of partially dissipative dark matter is a numerical test of the proposal in hydrodynamical simulations.  Such a simulation requires a prescription containing equations governing the evolution of energy densities and the number densities of each species in the dark sector, similar to simulations of baryons based on the radiative cooling rates. Our goal in this article is to provide such a cooling prescription for the dark sector as an input to numerical simulations, which could hopefully shed light on the feasibility of such a multi-component dark matter scenario and its variants.

Following Ref.~\cite{Fan:2013yva, Fan:2013tia}, we will consider the simplest partially dissipative dark matter scenario with the dissipative dark sector containing a dark electron (``coolant" $C$) and a dark proton (denoted as $X$), completely analogous to the ordinary baryonic sector. Both of them feel a dark $U(1)$ force mediated by a massless dark photon $\gamma_D$. The dark sector is assumed to be asymmetric, i.e. that there is a relative overabundance of dark electrons and protons compared to their anti-particles.

In the early Universe, $C$ and $X$ could become bound into dark atoms. Yet when galaxies form and dark matter falling into galaxies is shock-heated to the virial temperature, the dark atoms are ionized and $C$ and $X$ form a dark plasma distributed throughout the halo. The dark plasma then cools through Compton scattering of $C$ on the dark cosmic background photons and through dark bremsstrahlung. 
When the temperature drops to around the dark binding energy, dark recombination happens again. Further cooling is possible through recombination and ion-atom collisional processes. In short, cooling of the dissipative dark sector is basically parallel to that of the ordinary baryons. In this article, we will derive the energy loss rates of radiative cooling processes that transfer energy from the dark ions and dark atoms to the dark photon background. We will also present ionization and recombination coefficients (rates of the ionization and recombination processes) that change the fractions of dark ions and atoms. These rates can be incorporated into cosmological hydrodynamical simulations such as Illustris~\cite{Vogelsberger:2014kha, Vogelsberger:2014dza}, Eagle~\cite{Schaye:2014tpa} and Horizon-AGN~\cite{Dubois:2014lxa}.

The scenario we study resembles the traditional atomic dark matter models~\cite{Goldberg:1986nk, Mohapatra:2001sx, Kaplan:2009de, Feng:2009mn, Behbahani:2010xa, Cline:2013pca} but with two important differences: the dark ions and atoms collide to dissipate energy, and the dissipative sector is only a subdominant component of dark matter. This is in contrast with Refs~\cite{Foot:2014uba, Foot:2016wvj}, in which all of the dark matter is assumed to be dissipative.  

The derivations presented here are based on several approximations, which we justify. The goal is to present relatively simple and easily computable analytic equations with explicit dependence on the free parameters in the model, including dark electron mass $m_C$, dark proton mass $m_X$, dark fine-structure constant $\alpha_D$ and temperature $T$.
As a consistency check, we show that the equations yield results agreeing with numerical formulas from the literature on baryonic radiative cooling when the free parameters take the standard model values.

Except where otherwise noted, derivations are performed in natural units, in which $\hbar = c = k_B = \epsilon_0 = 1$, and final answers are presented in cgs units, following conventions in the astrophysics literature. Our notation is summarized in Table~\ref{tab:notation}.

\begin{table}[h]
  \centering
    \begin{tabular}{rl}
    \toprule
    \textbf{Symbol} & \textbf{Meaning} \\
    \midrule
    $X$ & dark proton\\
    $C$ & dark electron\\
    $H_D$ & dark hydrogen atom (bound state of $X$ and $C$) \\
    $\gamma_D$ & dark photon\\
    $n_C$ & dark electron number density \\
    $n_X$ & dark proton number density \\
    $n_{H_D}$ & dark atom number density \\
    $\sigma$ & cross section  \\
    $\alpha_D$ & dark fine-structure constant \\
    $T$   & temperature \\
    $K$   & kinetic energy of incident particle before collision \\
    $v$   & relative speed between the reactant particles \\
    $m_C$ & dark electron mass\\
    $m_X$ & dark proton mass \\
    Ry & dark Rydberg energy: $\frac{1}{2} \alpha_D^2 m_C$ \\
    $a_0$ & dark Bohr radius: $\frac{1}{\alpha_D m_C}$ \\
    $P$ & thermally averaged energy loss rate per unit volume\\
    $R$ & thermally averaged process rate per unit volume\\
    \bottomrule
    \end{tabular}%
      \caption{Summary of the notation used in paper.}
  \label{tab:notation}%
\end{table}%

The paper is organized as follows. 
In Section \ref{sec:processes} we derive or quote energy loss rates due to Compton scattering, bremsstrahlung, recombination, collisional ionization, and collisional excitation. For processes that change the number densities of different species, we also present the rates at which the processes occur. In Section~\ref{sec:heating}, we quote the rate of the reverse process, photoionization, that transfers energy from dark photons to dark atoms. 
In Section \ref{sec:approx}, we discuss the parameter space to which our results are applicable and justify the approximations used in our derivations. We conclude in Section~\ref{sec:conclusion}.

\section{Cooling Processes and Rates}
\label{sec:processes}

In this section, we calculate the rates at which the light dark ions, $C$, lose kinetic energy due to the following processes:
\begin{itemize}
\item Inverse Compton scattering: $C\gamma_D \to C\gamma_D$; 
\item Bremsstrahlung: $XC\to XC \gamma_D$; 
\item Recombination: $XC \to H_D\gamma_D$; 
\item Collisional ionization: $H_D C \to X C C$; 
\item Collisional excitation: $H_D C \to H_D^* C \to H_D C \gamma_D$; 
\end{itemize}
in which $H_D^*$ is an excited state of the dark atom. 
We also calculate the rates of the processes that change the number densities of different species. 

We assume that the dark plasma is optically thin, i.e., that dark photons emitted in cooling processes pass through it without being reabsorbed. We only consider the case with $m_X \gg m_C$. Thus the reduced mass of the dark atom can be approximated by $m_C$. We also assume that the dark electrons' speeds follow a Maxwell-Boltzmann distribution. Since the dark electron, $C$, moves more quickly than the heavier dark proton, $X$, the relative speed between the $C$ and $X$ (or $H_D$) can be approximated by $C$'s speed. 
We make several additional approximations in deriving different rates. All of the assumptions and approximations are discussed in more detail in Section \ref{sec:approx}.

Under these assumptions, if $\sigma_i(v)$ is the cross section of process $i$, given that the reactants collide at a relative speed $v$, the rate of process $i$ per volume is
\begin{equation}
R_i =n_A n_B \langle \sigma_i v \rangle,
\end{equation}
where $n_A$ and $n_B$ are the number densities of the reactants $A$ and $B$. The $\langle \cdots \rangle$ represents the thermal average. 
Similarly, if a collision at a relative speed $v$ leads to an energy loss of the incident particle, $E_\ell(v)$, the rate of energy loss per volume is
\begin{equation}
P_i = n_A n_B \langle E_\ell \sigma_i v \rangle.
\end{equation}
The thermal averages are performed over the Maxwell-Boltzmann distribution:
\begin{equation}
f(v) = \left(\frac{m_C}{2\pi T}\right)^{3/2} 4\pi v^2 e^{-\frac{m_Cv^2}{2T}}.
\end{equation}

\subsection{Inverse Compton Scattering}
\label{sec:compton}

In inverse Compton scattering, an energetic dark electron scatters off of a dark photon and transfers some of its energy to the photon. A simple way to compute the energy loss rate due to this process is to transform to the rest frame of the electron before the collision. In this frame, as long as the photon energy is much smaller than the dark electron mass, the process reduces to Thomson scattering. One can find the acceleration of the electron due to the incident electromagnetic radiation and then calculate the power that it radiates. The energy loss rate per volume in the nonrelativistic limit is~\cite{Peebles:1994xt}
\beq
P_{\rm Compton} &=& \frac{4(T-T_\gamma)}{m_C} \sigma_T n_C \frac{\pi^2}{15} T_\gamma^4, \nonumber \\
&=& \frac{n_C}{\mbox{cm}^{-3}} 1.9 \times 10^{-37} \mbox{erg/cm}^3\mbox{/s} \frac{T-T_\gamma}{1 \mbox{K}} \left(\frac{511 \mbox{keV}}{m_C}\right)^3 \left(\frac{\alpha_D}{10^{-2}}\right)^2\left(\frac{T_\gamma}{1 \mbox{K}}\right)^4,
\label{eq:compton}
\eeq
where $T_\gamma$ is the dark photon background temperature, which depends on the redshift $z$. Suppose that the current dark photon temperature is $T_\gamma^0 = T_\gamma(z=0) = T_{\rm CMB}/2 \approx 1.35$ K, then $T_\gamma(z)=(1+z)\times 1.35$ K. $\sigma_T$ is the dark Thomson scattering cross section:
\begin{equation}
\sigma_T = \frac{8\pi}{3} \left(\frac{\alpha_D}{m_C}\right)^2.
\end{equation}
Note that this process does not affect number densities of dark electrons. 

\subsection{Bremsstrahlung}
In thermal bremsstrahlung, dark electrons scatter off dark protons. As an electron is accelerated, it radiates away some of its initial energy. This process can be treated classically, provided that we restrict the impact parameter $b$ to 
\beq 
b > b_{\rm min}^{(1)} \equiv \frac{2\pi}{m_C v},
\eeq
the limit set by the uncertainty principle. To the lowest order, we can assume that the electron's trajectory is not modified by the loss of energy and is approximately a straight line, and we can neglect all forces other than Coulomb attraction, provided that 
\beq
b > b_{\rm min}^{(2)} \equiv \frac{4\alpha_D}{\pi m_C v^2},
\eeq
which can be obtained by stipulating that the change in velocity due to the acceleration normal to the trajectory is much smaller than the velocity. Strictly speaking, only when $b_{\rm min}^{(2)} \gg b_{\rm min}^{(1)}$ or, equivalently, the electron kinetic energy is much smaller than the Ryderberg energy of the dark atom, can the classical description together with the straight-line approximation be applied. However, the formula obtained still has the correct parametric dependence and the correction from a full quantum treatment will only modify the classical result by a free-free Gaunt factor $g_{\rm ff}$. 

In the nonrelativistic limit, the energy loss rate per volume due to bremsstrahlung is \cite{draine-2011}
\begin{equation}
\begin{split}
P_{\rm brems} &= n_X n_C \frac{16}{3} \sqrt{\frac{2\pi}{3}} \frac{\alpha_D^3}{m_C^2} \sqrt{m_CT} \bar{g}_{\rm ff}\\
&=  \frac{n_C}{{\rm cm}^{-3}} \frac{n_X}{{\rm cm}^{-3}} 3.7 \times 10^{-27}~{\rm erg/cm^3/s}\left(\frac{511~{\rm keV}}{m_C}\right)^{3/2} \left(\frac{\alpha_D}{10^{-2}}\right)^3 \left(\frac{T}{1~{\rm K}}\right)^{1/2} \bar{g}_{\rm ff},
\end{split}
\label{eq:brem}
\end{equation}
where $\bar{g}_{\rm ff}$ is the thermally averaged free-free Gaunt factor.
As a first order approximation we take $\bar{g}_{\rm ff}$ to be 1, which will be justified in Section 3. As is the case for inverse Compton scattering, bremsstrahlung does not affect number densities of dark electrons, so we do not have to compute the process rate.

\subsection{Recombination}
\label{sec:recombination}

In recombination, a free dark electron and a free dark proton recombine to form a dark hydrogen atom, radiating away a fraction of the incident electron's kinetic energy. However, the temperature, which is proportional to the average kinetic energy of the free electrons, does not change. We will estimate the rate, $R_{\rm rec}$, at which this process occurs, which is necessary to keep track of the number densities of ions and atoms, as well as the kinetic energy loss rate.

It is standard to derive the recombination cross section from the photoionization cross section (e.g. Ref.~\cite{karzas_latter_1961}) using the Milne relation, which is a detailed balancing relation \cite{rybicki_lightman_1979}. Below we will discuss a simper and more intuitive method following Ref.~\cite{gould_thakur_1970}. The key point of this method is that recombination can be treated as a special case of bremsstrahlung in which the electron radiates enough energy to become bound to the proton. This argument is heuristic, as the bremsstrahlung spectrum is derived by assuming that the electron trajectory is nearly unmodified by the radiation reaction force, whereas in this case it is heavily modified as the electron is captured. Nevertheless, the result obtained has the correct parametric dependence and only differs from the quantum result by a bound-free Gaunt factor $g_{\rm bf}$.

The differential cross section for emitting a dark photon with frequency $\omega$ through bremsstrahlung off of a dark proton is
\begin{equation}
d\sigma_{\rm brem} = \frac{16 \pi \alpha_D^3}{3^{3/2} m_C^2 v^2 \omega}  d\omega.
\label{eq:diffbrem}
\end{equation}
If the dark electron is captured by the proton and recombines into a bound state with principal quantum number $n$, the frequency of the dark photon is given by
\begin{equation}
\omega_n = K + \frac{\alpha_D^2m_C }{2 n^2},
\end{equation}
where $K = m_C v^2/2$ is the initial kinetic energy of the dark electron. This leads to
\begin{equation}
|\Delta \omega| = \frac{\alpha_D^2m_C \Delta n}{n^3}. \nonumber
\end{equation}
Note that $\Delta n$ is not really a differential but is equal to unity. Plugging this into Eq.~\ref{eq:diffbrem} and replacing $d\omega$ by $\frac{\Delta \omega}{\Delta n} \Delta n$, the cross section for recombining to an energy level $n$ is
\begin{equation}
\sigma_{{\rm rec},n} = \frac{32 \pi \alpha_D^5}{3^{3/2} m_C^2v^2 n^3(v^2 + \frac{\alpha_D^2}{n^2} )}.
\end{equation}
A full quantum calculation gives the same result multiplied by a Gaunt factor $g_{\rm bf}$. The Gaunt factor depends on $n$ and $K/{\rm Ry} = v^2/\alpha_D^2$ where the Rydberg energy is Ry=$\frac{\alpha_D^2m_C}{2}$. 

It is well known that during the recombination epoch of ordinary baryons before galaxy formation, electrons and protons dominantly recombine into the $2s$ or $2p$ states~\cite{1968ApJ}. Atoms in the $2p$ state decay to the ground state by emitting a Lyman-$\alpha$ photon while atoms in the $2s$ states decay to the ground state by simultaneous emission of two photons. The direct recombination into the ground state is negligible since the density of hydrogen nuclei is very high $\sim$ 400 cm$^{-3}$ (at $z \approx 1300$) and the emitted photons are absorbed very quickly. Similarly, because the Lyman-$\alpha$ photons usually get re-absorbed, the rare $2s\to 1s$ transition becomes important since neither of the two photons from the $2s$ decay has enough energy to excite the atom. The recombination considered here happens in galaxies with much lower atomic density and the gas is optically thin (more details can be found in Section~\ref{sec:opticallythin}). Thus we include recombination to the ground state as well as to the excited states.  

\subsubsection{Recombination Rate}
The thermally averaged recombination rate, summed over all energy levels, is 
\begin{equation}
\label{eq:recombRate}
\begin{split}
\frac{R_{\rm rec}}{n_Xn_C} &\equiv \sum_{n=1}^\infty \langle \sigma_{{\rm rec},n} v  \rangle = \frac{2^{11/2} \pi^{1/2}  \alpha_D^5}{3^{3/2} m_C^{1/2} T^{3/2}} \int_0^\infty \sum_{n=1}^\infty \frac{u e^{-u^2} }{u^2n^3 + y^2 n} g_{\rm bf}(n, u^2/y^2) du, \quad y^2 \equiv \frac{\rm Ry}{T} = \frac{m_c\alpha_D^2}{2T} \\
&\approx \begin{cases} \frac{2^{9/2}\pi^{1/2}\alpha_D^3}{3^{3/2}m_C^{3/2}T^{1/2}} \left[1.744 + \log\left(y^2\right) + \frac{1}{6y^2} \right] \quad  \quad \quad \quad \quad \quad \quad \quad \quad \; y \gg 1  \\
=8.4\times 10^{-14} \frac{{\rm cm}^3}{{\rm s}} \left(\frac{\alpha_D}{10^{-2}}\right)^3 \left(\frac{511\,{\rm keV}}{m_C}\right)^{3/2}\left(\frac{10^5\,{\rm K}}{T}\right)^{1/2}\left[1.744 + \log y^2 + \frac{1}{6y^2} \right],  
\\	
\frac{2^{5/2} \pi^{1/2}  \alpha_D^5}{3^{5/2} m_C^{1/2} T^{3/2}} \left[ -4.66 - 15 \log y^2 + y^2(5.42- 14 \log y^2)\right] \quad \quad y \ll 1 \\
= 1.3\times 10^{-15} \frac{{\rm cm}^3}{{\rm s}} \left(\frac{\alpha_D}{10^{-2}}\right)^5 \left(\frac{511\,{\rm keV}}{m_C}\right)^{1/2}\left(\frac{10^6\,{\rm K}}{T}\right)^{3/2}\left[ -4.66 - 15 \log y^2 + y^2(5.42- 14 \log y^2)\right], 
 \end{cases}
\end{split}
\end{equation}
where in second lines, we have set $g_{\rm bf}$ to be 1, approximated the sum over $n$ using the Euler-Maclaurin formula and expanded in the indicated limits. The result for $y \gg 1$ was also obtained in Ref.~\cite{gould_thakur_1970}.  

\begin{figure}[h]
\centering
\includegraphics[width=0.6\textwidth]{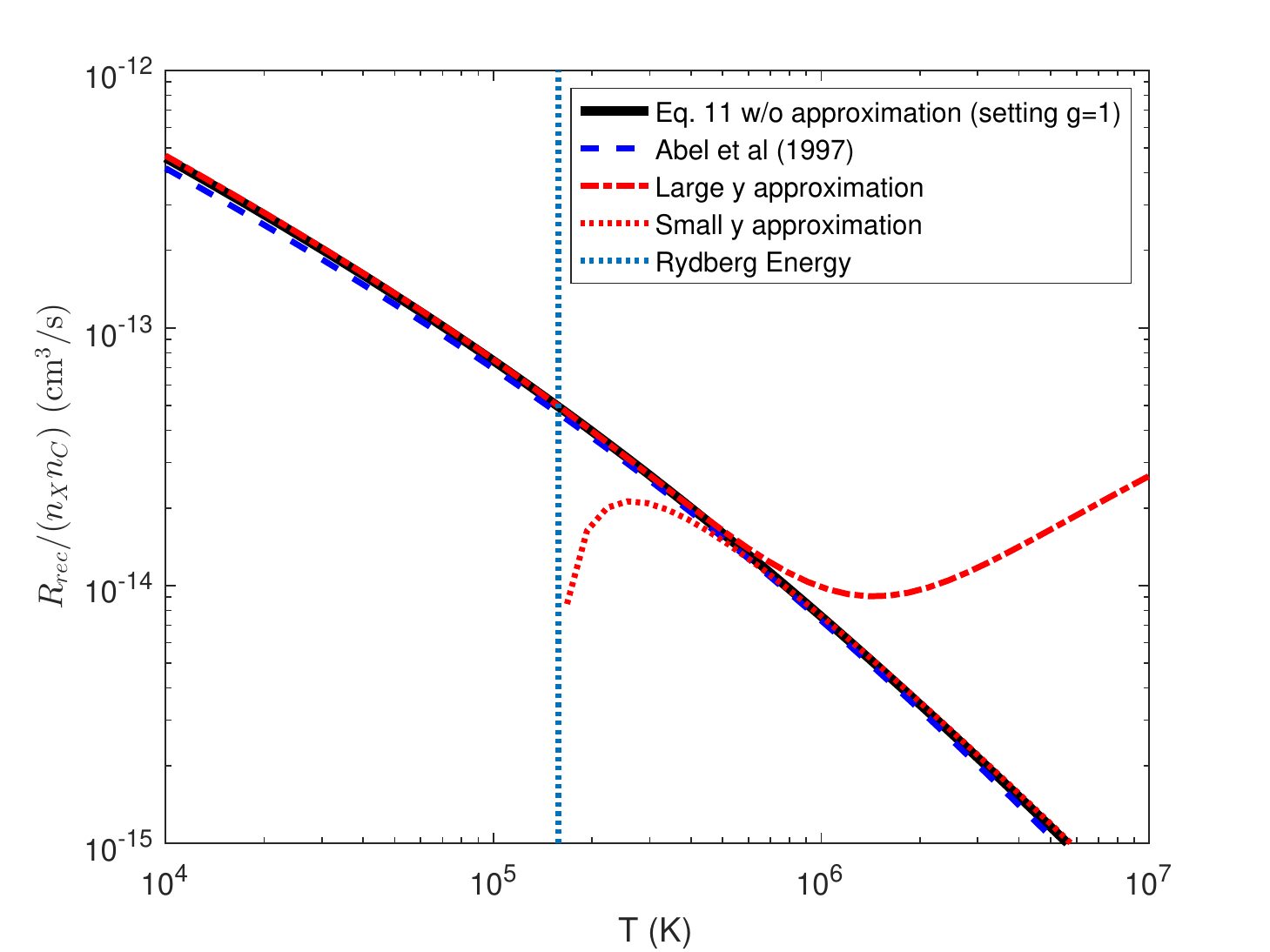}
\caption{Radiative recombination rates as a function of $T$ using Eq.~\ref{eq:recombRate}, either the full sum formula (black, solid) or different limits (large $y$: red dashed-dot; small $y$: red dotted), compared to the fit (blue dashed) of Ref.~\cite{Abel:1996kh} (which is a fit to the analytic formula given by Ref.~\cite{Ferland1992}), setting the parameters to the standard model values. The vertical dashed line corresponds to $T =$ Ry around which dark recombination happens and the numbers of ions start to decrease.  }
\label{fig:recombinationComparison}
\end{figure}

In Figure \ref{fig:recombinationComparison}, we compare the full formula (first line of Eq.~\ref{eq:recombRate}), the different limits (rest of Eq.~\ref{eq:recombRate}), and the result quoted in Ref.~\cite{Abel:1996kh}, assuming the standard model parameters. The figure demonstrates that, given the standard model values, the full formula as a sum over all energy levels in Eq.~\ref{eq:recombRate} matches the numerical formula in Ref.~\cite{Abel:1996kh} very well. The analytic formulas in the large and small $y$ limits agree almost exactly with the full formula in their valid regimes, and the two limits merge around $T\sim$ Ry.

\subsubsection{Energy Loss Rate}
\begin{figure}[h]
\centering
\includegraphics[width=0.65\textwidth]{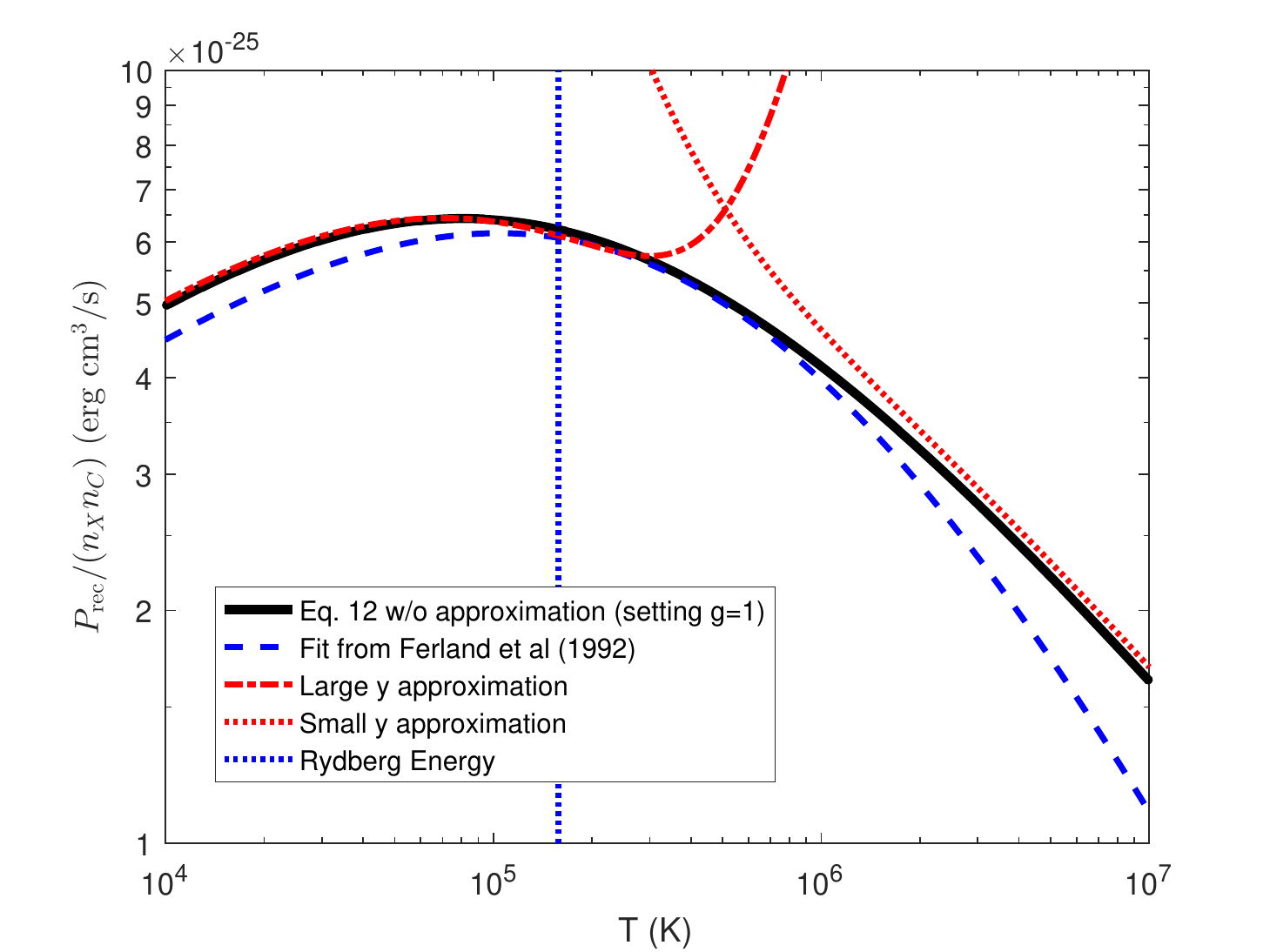}
\caption{Energy loss rate (normalized by $n_Cn_X$) due to radiative recombination as a function of $T$, using Eq.~\ref{eq:rateofenergylossrec}, either the full sum formula (black, solid) or different limits (large $y$: red dashed-dot; small $y$: red dotted), compared to the fit formula in Ref.~\cite{Ferland1992} (blue dashed), setting the parameters to the standard model values. As in Fig. \ref{fig:recombinationComparison}, the vertical dashed line corresponds to $T =$ Ry around which dark recombination happens.}
\label{fig:recombinationPowerComparison}
\end{figure}

The thermally averaged (kinetic) energy lost rate per volume, divided by $n_C$ and $n_X$, is given by
\begin{equation}
\label{eq:rateofenergylossrec}
\begin{split}
\frac{P_{\rm rec}}{n_X n_C} &=\sum_{n=1}^\infty \left\langle \left(\frac{1}{2} m_C v^2 \right)\sigma_{{\rm rec},n} v \right\rangle \\
&= \frac{2^{11/2} \pi^{1/2}  \alpha_D^5 }{3^{3/2}  m_C^{1/2}T^{1/2}}  \int_0^\infty \sum_{n=1}^\infty \frac{u^3 e^{-u^2}}{n^3u^2 + y^2n} g_{\rm bf}(n, u^2/y^2)  du \quad y^2 \equiv \frac{m_c\alpha_D^2}{2T}\\
&\approx \begin{cases} \frac{2^{9/2} \pi^{1/2}  \alpha_D^3 T^{1/2} }{3^{3/2}  m_C^{3/2}} \left[ 0.74 + \log y^2 + \frac{1}{3y^2} \right] \\
= 1.2 \times 10^{-24} \frac{\rm erg \cdot{\rm cm}^3}{{\rm s}}\left(\frac{\alpha_D}{10^{-2}}\right)^3 \left(\frac{511\,{\rm keV}}{m_C}\right)^{3/2}\left(\frac{T}{10^5\,{\rm K}}\right)^{1/2} \left[ 0.74 + \log y^2 + \frac{1}{3y^2} \right]	&y\gg1 \\ 	
\frac{2^{5/2} \pi^{1/2}  \alpha_D^5 }{3^{3/2} m_C^{1/2}T^{1/2} } 	\left[5 + y^2(2.860 + \frac{14}{3} \log y^2)\right] \\
=5.4 \times 10^{-25} \frac{\rm erg \cdot{\rm cm}^3}{{\rm s}}\left(\frac{\alpha_D}{10^{-2}}\right)^5 \left(\frac{511\,{\rm keV}}{m_C}\right)^{1/2}\left(\frac{10^6\,{\rm K}}{T}\right)^{1/2} \left[5 + y^2(2.860 + \frac{14}{3} \log y^2)\right]&y\ll1	,\end{cases}
\end{split}
\end{equation}
where $n_X, n_C$ are in units of 1/cm$^3$. 
In deriving the two limits, we approximate the sum using the Euler-Maclaurin formula as before. The rates computed using different lines of Eq. \ref{eq:rateofenergylossrec} with the standard model parameters are shown in Figure \ref{fig:recombinationPowerComparison}. Again there is good agreement between our results and the numerical formula in Ref.~\cite{Ferland1992} within 50\% in the entire $T$ range. 

\subsection{Collisional Ionization}
\label{sec:collisionalIonization}
In collisional ionization, a charged particle collides with a dark hydrogen atom and ionizes it. The charged particle can be either a dark proton or a dark electron. Here we consider only dark electrons since the rate of collisional ionization due to proton impact is small (see discussion in Section \ref{sec:neglectProtons}). We use the binary encounter approximation, in which the dark hydrogen nucleus is ignored and the bound electron is treated as a free electron at rest. The atom is considered ionized if the final kinetic energy of the bound electron is greater than the binding energy. We only consider ionization from the ground state. 

Using the binary encounter approximation, the cross section can be calculated classically (it was first obtained by Thomson in 1912, cf Ref.~\cite{peterkop_1977}):
\begin{equation}
\sigma_{\rm ion, binary} = \frac{4\pi}{m_C^2v^2} \left( 1 - \frac{\alpha_D^2}{v^2} \right) = 4 \pi a_0^2\; x (1-x), \quad {\rm with} \quad x = \frac{\alpha_D^2}{v^2} = \frac{\rm Ry}{K},
\label{eq:BE}
\end{equation}
where $a_0=1/(\alpha_D m_C)$ is the Bohr radius.
This approximation can be further improved. For instance, Burgess accounts for an exchange interaction (due to the fact that the scattered and ejected electrons are indistinguishable) and for the fact that the incident electron gains a kinetic energy (e.g. of the order the binding energy) due to the nucleus prior to the collision~\cite{burgess_1964}. More recently, Kim and Rudd developed the binary-encounter-Bethe (BEB) model which combines the binary encounter approximation and the Bethe theory for fast ($v \gg \alpha_D$) incident electrons~\cite{Kim1994}. The total ionization cross section based on the BEB model is given by
\beq
\sigma_{\rm ion, BEB} = 4\pi a_0^2 \;\frac{x}{1+2x} \left(1-x - \frac{1-x^2}{2} \ln x +\frac{x \log x}{1+x}\right),\quad {\rm with} \quad x = \frac{\alpha_D^2}{v^2} = \frac{\rm Ry}{K}.
\label{eq:improvedion}
\eeq
We set the differential oscillator strength in the BEB model to one to get the equation above. This is generally true to the leading order except for resonances, which are unimportant for our discussion. 
Both cross sections are of the same order of magnitude but differ by an order one number numerically.  For the purpose of simulation, either formula could be used since an order one number will not alter the result significantly. 

The ionization rate is then
\beq
\frac{R_{\rm ion}}{n_C n_{H_D}} &=&  \langle \sigma_{\rm ion} v\rangle  \nonumber \\
&=&\frac{2^{7/2}\sqrt{\pi}}{m_C^{3/2} T^{1/2}} \bar{f}(y^2), \quad y^2=\frac{\alpha_D^2 m_C}{2T} \nonumber \\
&=& 2.2 \times 10^{-7} \,{\rm cm}^3/{\rm s} \left( \frac{511 \, {\rm keV}}{m_C}\right)^{3/2} \left( \frac{10^5 \,{\rm K}}{T}\right)^{1/2}   \bar{f}(y^2),
\label{eq:ionrate}\eeq
where $n_C$ and $n_{H_D}$ are in units of $1/$cm$^3$ and the function $\bar{f}(y^2)$ is defined as 
\begin{equation}
\bar{f} (y^2)=\begin{cases}
\int_{y}^\infty \left(1-\frac{y^2}{u^2}\right)u e^{-u^2} du,   & {\rm for} \; \sigma_{\rm ion, binary} \\
\int_{y}^\infty\frac{ue^{-u^2}}{1+2y^2/u^2} \left(1-\frac{y^2}{u^2} - \frac{1-\left(\frac{y^2}{u^2}\right)^2}{2} \ln \left(\frac{y^2}{u^2}\right) +\frac{\frac{y^2}{u^2} \log \frac{y^2}{u^2}}{1+\frac{y^2}{u^2}}\right) du   & {\rm for} \; \sigma_{\rm ion, BEB}
\end{cases}
\end{equation}

The rate of electron kinetic energy loss per volume, normalized by $n_Cn_{H_D}$, is then 
 \beq
 \frac{P_{\rm ion}}{n_C n_{H_D}} &=& \frac{1}{2}\alpha_D^2m_C  \frac{R_{\rm ion}}{n_Cn_{H_D}} \nonumber \\
 &=& 9 \times 10^{-18} \frac{\rm erg \cdot{\rm cm}^3}{{\rm s}}\left(\frac{\alpha_D}{10^{-2}}\right)^2 \left( \frac{511 \, {\rm keV}}{m_C}\right)^{1/2} \left( \frac{10^5 \,{\rm K}}{T}\right)^{1/2}   \bar{f}(y),
 \label{eq:ionenergyrate}
\eeq
where $n_C$ and $n_{H_D}$ are in units of $1/$cm$^3$.

In Figure \ref{fig:collisionalIonizationComparison}, we compare our energy loss rates based on either the cross section in Eq.~\ref{eq:BE} or Eq.~\ref{eq:improvedion} with the result quoted in Ref.~\cite{Abel:1996kh}, assuming standard model input values. The binary encounter approximation and Ref.~\cite{Abel:1996kh} agree within a factor of 2-3, while the BEB model and Ref.~\cite{Abel:1996kh} agree within 20\%.

\begin{figure}[H]
\centering
\includegraphics[width=0.65\textwidth]{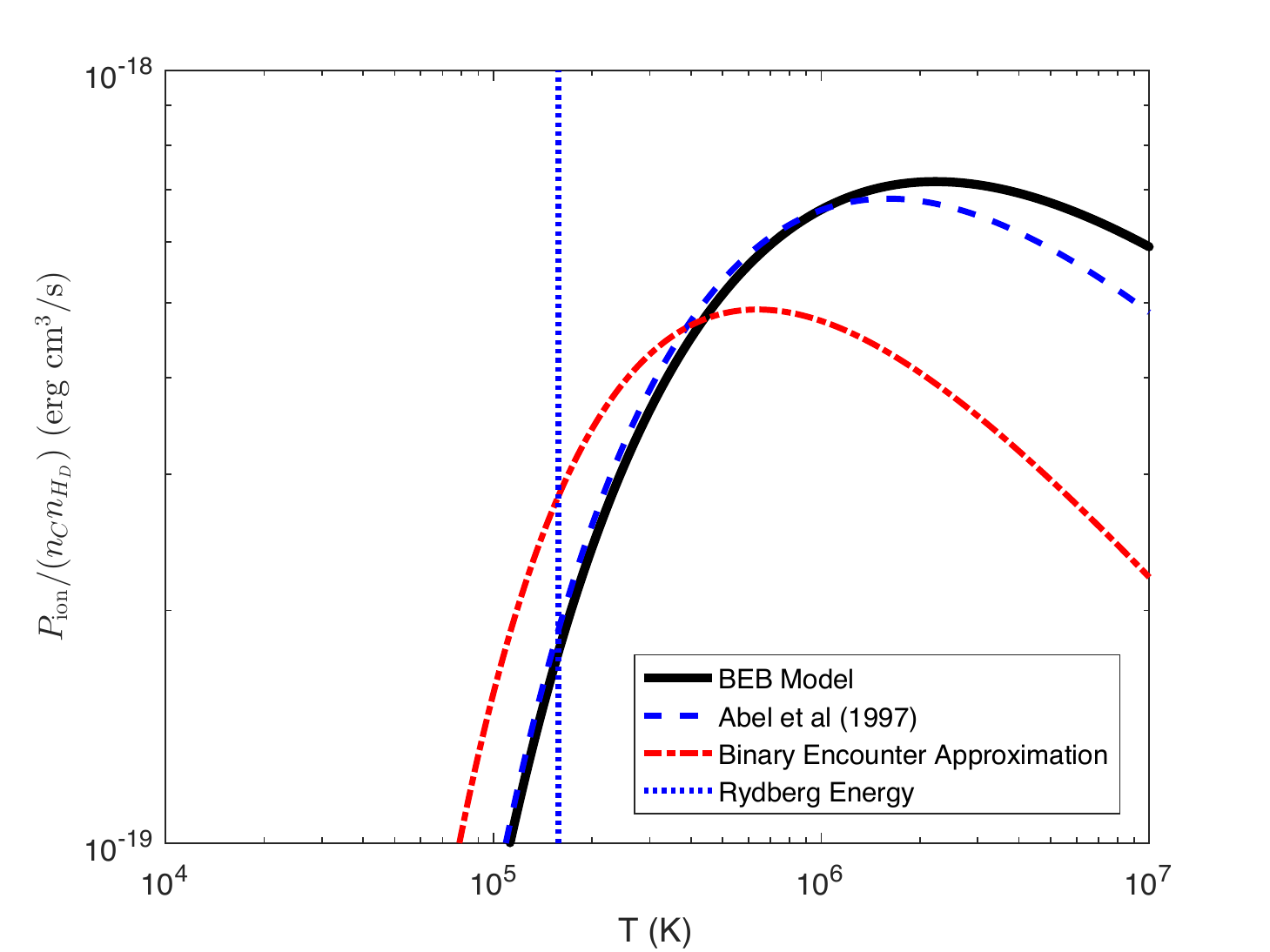}
\caption{Kinetic energy loss rate as a function of temperature due to collisional ionization, divided by $n_C$ and $n_{H_D}$, using the standard model parameters. The results based on Eq.~\ref{eq:BE} (red, dash-dotted) and Eq.~\ref{eq:improvedion} (black, solid) agree with the formula quoted in Ref.~\cite{Abel:1996kh} (blue, dashed). The vertical dotted line corresponds to $T=$ Ry around which the collisional processes are effective. }
\label{fig:collisionalIonizationComparison}
\end{figure}

\subsection{Collisional Excitation}
\label{sec:collisionalExcitation}
Collisional excitation is the inelastic collision of a free dark electron with a dark hydrogen atom. The atom is excited to a higher energy level and the incident electron loses energy. The excited state then decays back to the lower energy level and releases the energy as dark photons escape the dark plasma. We only consider the transition from the ground state. 

We compute the rates using the plane-wave Born approximation. The Hamiltonian for collisional excitation in the position basis is
\begin{equation}
H = -\frac{1}{2m_C} \nabla^2_1 - \frac{1}{2m_C} \nabla^2_2- \frac{\alpha_D}{r_2}  - \frac{\alpha_D}{r_1} + \frac{\alpha_D}{|\mathbf{r_1} - \mathbf{r_2}|},
\end{equation}
where $\bf r_1$ is the position of the free electron, and $\bf r_2$ is the position of the bound electron. In the Born approximation, $H_1 \equiv -\frac{\alpha_D}{r_1} + \frac{\alpha_D}{|\mathbf{r_1} - \mathbf{r_2}|}$ is treated as a small perturbation on the rest of the Hamiltonian (cf. Ref.~\cite{schiff_1968}). The scattering amplitude ${\cal M}$ is then proportional to $\langle f|H_1|i\rangle$, where the initial and final states are products of free particle wavefunctions and hydrogen wavefunctions. After integrating over $\bf r_1$, we find that
\begin{equation}
{\cal M}(\Omega) = -(2\pi)^2 m_C \sqrt{\frac{k'}{k}}\langle f | H_1|i\rangle = \frac{2\alpha_D m_C}{ q^2}\sqrt{\frac{k'}{k}} \int d^3 \mathbf{r_2} \psi^*_{n'l'm'}(\mathbf{r_2}) \psi_{nlm}(\mathbf{r_2}) e^{-i\mathbf{q}\cdot \mathbf{r_2}},
\end{equation}
where $\bf k$ and $\bf k'$ are the initial and final momentum of the free electron, respectively, and $\mathbf{q} = \mathbf{k'} - \mathbf{k}$ is the momentum transfer. By energy conservation, the difference in magnitude of $\bf k$ and $\bf k'$ is determined by the change in energy level of the hydrogen atom from $n$ to $n'$. For the particular case of a $1s\to2p$ transition, we have
\begin{equation}
\frac{d\sigma_{1s\to2p, {\rm Born}}}{d\Omega} = |{\cal{M}}(\Omega)|^2 = \frac{2^{17}}{3^{10}q^2 \left(1+\frac{4}{9}\left(\frac{q}{\alpha_Dm_C}\right)^2\right)^6}  \frac{k'}{k}.
\end{equation}
The total cross section can then be obtained by changing variables from $d(\cos\theta)$ to $dq=-\frac{kk'}{q} d(\cos\theta)$ and integrating over $dq$:
\begin{equation}
\label{eq:highEnergyLimit}
\begin{split}
\sigma_{1s\to2p, {\rm Born}} &= \frac{2\pi}{kk'} \int_{q_{min}}^{q_{max}} \frac{d\sigma}{d\Omega}qdq = \frac{2^{18}}{3^{10}}\left(\frac{\pi}{m_C^2 v^2}\right) \int_{(k-k^\prime)a_0}^{(k+k^\prime)a_0} \frac{dx}{x} \frac{1}{ \left(1+4x^2/9\right)^6} \\
&\approx \frac{2^{18}}{3^{10}}\left(\frac{\pi}{m_C^2 v^2}\right) \log\left(\frac{4v}{\alpha_D}\right)=4\pi a_0^2 \left(\frac{2^{15}}{3^{10}}\right) \frac{\rm Ry}{K} \log\left(\frac{16K}{\rm Ry}\right), \quad v \gg \alpha_D (K \gg {\rm Ry}),
\end{split}
\end{equation}
where in the second line, we take the limit $v \gg \alpha_D$ or equivalently, $K \gg$ Ry. The logarithm originates from $1/q^2$ in the differential cross section. For ordinary electron-impact collisional excitation, the leading-order Born approximation gives a result which agrees with the experimental data within a factor of 2. 
Ref.~\cite{kim_2001} demonstrates that an empirical scaling called BE scaling improves this result to match the data better, especially at lower incident kinetic energy ($v \sim \alpha_D$). It is defined as
\begin{equation}
\sigma_{nl \to n^\prime l^\prime, {\rm BE}} = \sigma_{ nl \to n^\prime l^\prime, {\rm Born}} \left(\frac{K}{K+{\rm Ry}+\Delta E}\right),
\label{eq:BEscaling}
\end{equation}
where $\Delta E$ is the excitation energy.\footnote{In Ref.~\cite{kim_2001}, {\rm Ry} is denoted by $B$ and $\Delta E$ is denoted by $E$ so it is called BE scaling.} We discuss the reasoning behind BE scaling in Section~\ref{sec:approx}. We have checked (not shown here) that the cross section obtained by integrating the first line of Equation \ref{eq:highEnergyLimit} numerically and then rescaled by Eq.~\ref{eq:BEscaling} matches the result in Ref.~\cite{kim_2001}. 

Finally, the energy loss rate (divided by the electron and atom number densities) is
\begin{equation}
\label{eq:1s2pPower}
\begin{split}
\frac{P_{1s\to2p}}{n_C n_{H_{D; 1s}}} &= \frac{3}{8} m_C \alpha_D^2 \langle \sigma_{ 1s\to 2p} v \rangle\\
&=7.4 \times 10^{-18}{{\rm erg} \cdot {\rm cm}}^3/{\rm s} \left(\frac{\alpha_D}{10^{-2}}\right)^2\sqrt{\frac{511{\rm keV}}{m_C}} \sqrt{\frac{10^5{\rm K}}{T}}g\left(\frac{m\alpha_D^2}{2T}\right) \\
{\rm where} \; g(y^2)&=  \int_{\frac{\sqrt{3}}{2}y}^\infty du \frac{u e^{-u^2}}{1+\frac{7y^2}{4u^2}} \int_{x_-}^{x_+} \frac{dx}{x} \frac{1}{ \left(1+\frac{4x^2}{9}\right)^6}, \; {\rm with} \quad x_\pm = \frac{u}{y} \left(1\pm \sqrt{1-\frac{3}{4}\frac{y^2}{u^2}}\right)  \\
&\approx \int du \frac{u e^{-u^2}}{1+\frac{7y^2}{4u^2}} \left(\log\left(\frac{4u}{y}\right)+\cdots \right)
\end{split}
\end{equation}
The lower bound in the integration of $u$ comes from the requirement that the kinetic energy of the incident electron has to be larger than 3/4 Ry to trigger the transition. In the second line of $g(y^2)$, we keep the leading term in the expansion at large $u/y$.  
 The results presented here use the BE rescaled Born cross section. 
 
 \begin{figure}[h]
\centering
\includegraphics[width=0.6\textwidth]{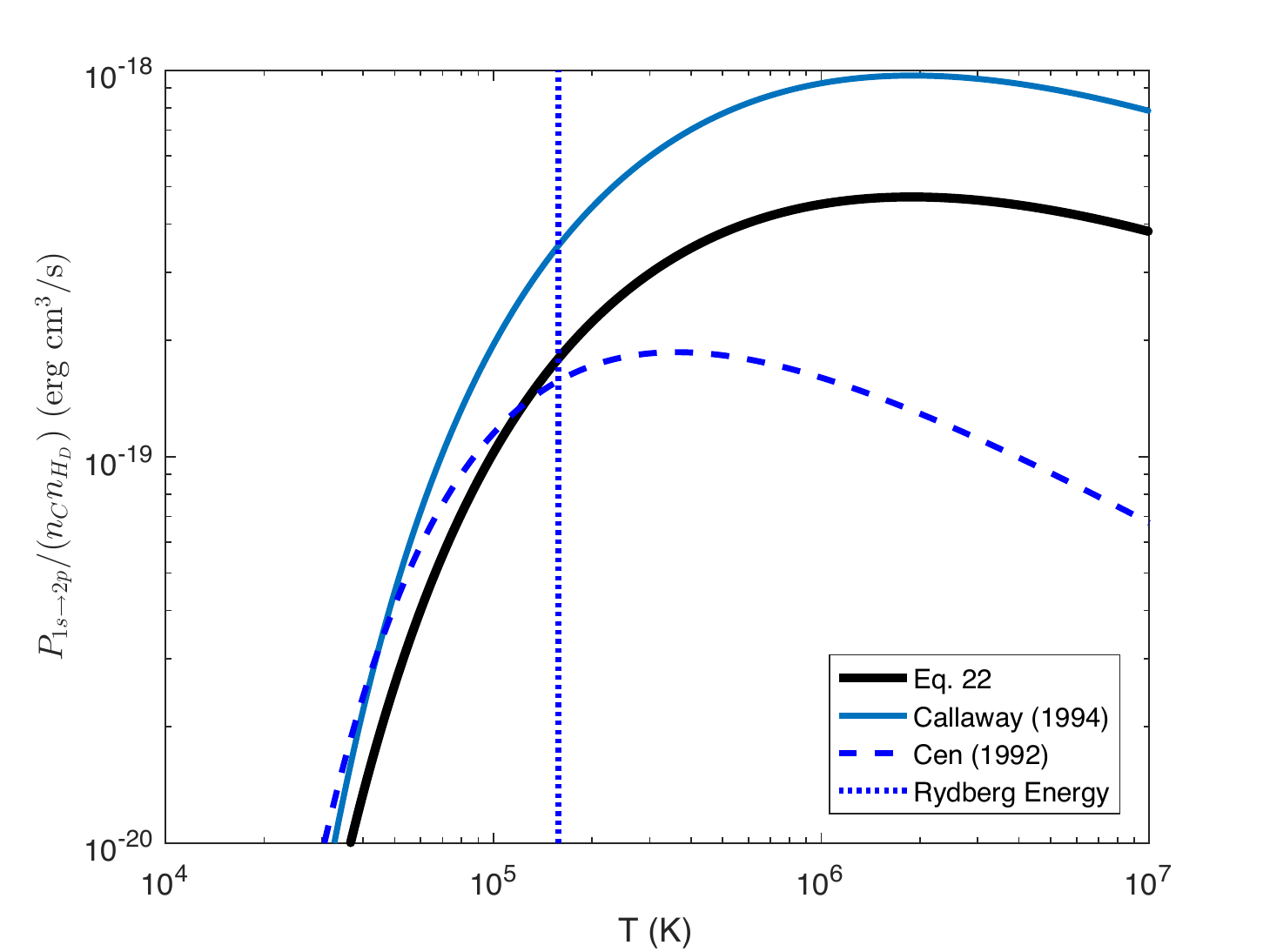}
\caption{Rates of energy loss due to collisional excitation as a function of temperature, using the standard model parameters. The result based on Eq.~\ref{eq:1s2pPower} (black, solid) agrees reasonably well with the fit in Ref.~\cite{CALLAWAY19949} (blue, solid) as well as the result quoted in Ref.~\cite{Cen:1992zk} (blue, dashed). The vertical dotted line corresponds to $T=$ Ry around which the collisional processes are effective. }
\label{fig:collisionalIonizationComparison}
\end{figure}

Let's compare this to the $1s \to 2s$ transition. Following the same procedure, we find that the total cross section for the $1s \to 2s$ transition is
\begin{equation}
\begin{split}
\sigma_{1s\to2s, {\rm Born}} &= \frac{2\pi}{kk'} \int_{q_{min}}^{q_{max}} \frac{d\sigma}{d\Omega}qdq = \frac{2^{20}}{3^{12}}\left(\frac{\pi}{m_C^2 v^2}\right) \int_{(k-k^\prime)a_0}^{(k+k^\prime)a_0} dx  \frac{x}{ \left(1+4x^2/9\right)^6} \\
&\approx \frac{2^{17}}{3^{10}\times5}\left(\frac{\pi}{m_C^2 v^2}\right)=4\pi a_0^2 \left(\frac{2^{15}}{3^{10}\times 5}\right) \frac{\rm Ry}{K}, \quad v \gg \alpha_D (K \gg {\rm Ry}).
\end{split}
\end{equation}
The radiated power is given by
\begin{equation}
\label{eq:1s2sPower}
\begin{split}
\frac{P_{1s\to2s}}{n_C n_{H_{D;1s}}} &= \frac{3}{8} m_C \alpha_D^2 \langle \sigma_{ 1s\to 2s} v \rangle\\
&=3.3 \times 10^{-18}{\rm erg \cdot \rm cm}^3/{\rm s} \left(\frac{\alpha_D}{10^{-2}}\right)^2\sqrt{\frac{511{\rm keV}}{m_C}} \sqrt{\frac{10^5{\rm K}}{T}}h\left(\frac{m\alpha_D^2}{2T}\right) \\
{\rm where} \;h(y^2)&=  \int_{\frac{\sqrt{3}}{2}y}^\infty du \frac{u e^{-u^2}}{1+\frac{7y^2}{4u^2}} \int_{x_-}^{x_+} dx \frac{x}{ \left(1+\frac{4x^2}{9}\right)^6}, \; {\rm with} \quad x_\pm = \frac{u}{y} \left(1\pm \sqrt{1-\frac{3}{4}\frac{y^2}{u^2}}\right)  \\
&\approx \int du \frac{u e^{-u^2}}{1+\frac{7y^2}{4u^2}} \left(\frac{9}{40} + \cdots\right).
\end{split}
\end{equation}
Comparing Eq.~\ref{eq:1s2pPower} and Eq.~\ref{eq:1s2sPower}, one can see that $1s\to2p$ dominates over $1s\to2s$ by at least a factor of 10. This is because the $1s \to 2p$ differential cross section peaks at small angles and the total scattering decreases less rapidly with increasing energy at high energies. 
Aiming to get the energy loss rate correct within an order of magnitude, we will ignore the $1s\to2s$ transition as well as all other transitions from the ground state to excited states except $1s\to2p$.

Using the standard model values of the input parameters, we plot the energy loss rate as a function of temperature based on Eq.~\ref{eq:1s2pPower} in Figure \ref{fig:collisionalIonizationComparison} and compare it to the results in Ref.~\cite{CALLAWAY19949} and Ref.~\cite{Cen:1992zk}. 
All three results agree with each other within an order of magnitude. Our result agrees with that in Ref.~\cite{CALLAWAY19949} within a factor of 2, which is better than the agreement with the older result quoted in Ref.~\cite{Cen:1992zk}.

\subsection{Summary}
\begin{figure}[h]
\centering
\includegraphics[width=0.65\textwidth]{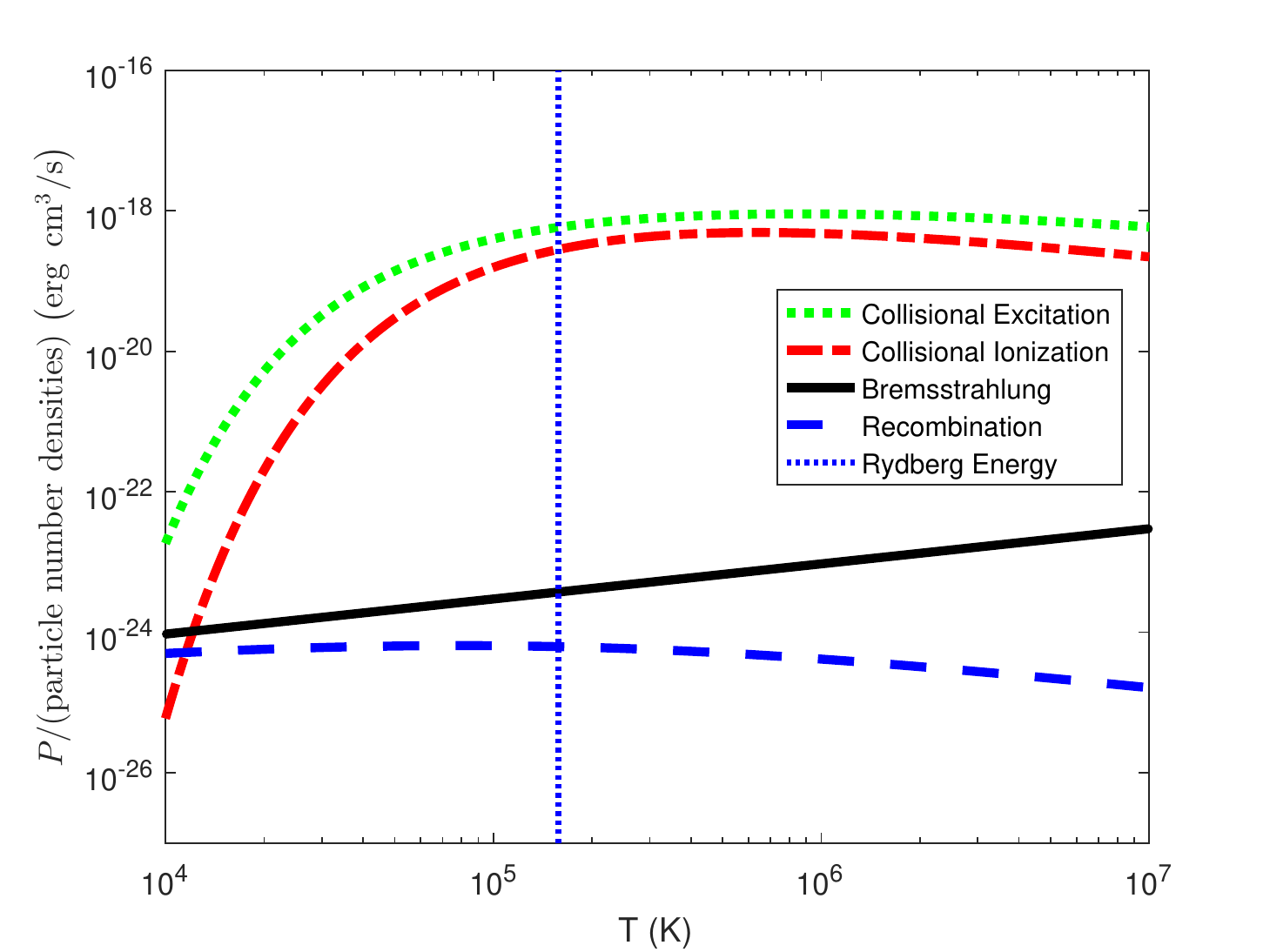}
\caption{ Comparison of the energy loss rates due to indicated processes as a function of temperature, assuming the standard model input values. Note that the relevant number densities are not the same for each process. Namely, the collisional ionization and excitation rates are proportional to $n_C n_{H_D}$, whereas the bremsstrahlung and recombination rates are proportional to $n_C n_X$. }
\label{plot:com}
\end{figure}

In Figure \ref{plot:com}, we present energy loss rates due to the processes studied in this section as a function of temperature, assuming the standard model input values. Note that the energy loss rates we present are normalized by the number densities of reactant particles. Even though energy loss rates due to the collisional processes are greatest at $T \gg$ Ry, there are very few dark atoms at such high temperatures, and so these processes are, in fact, suppressed compared to ionic processes such as bremsstrahlung. For ordinary baryons, simulations (e.g., Ref~\cite{Katz:1995up}) show that bremsstrahlung (free-free emission) dominates at $T > 10^6$K. At temperatures between $10^{4.3}$K and $10^5$K (1.7 eV - 8.6 eV), collisional processes are the dominant sources of cooling. Below $10^4$K, the gas is entirely neutral and cooling rate is essentially zero. 

\section{Heating Process: Photoionization}
\label{sec:heating}

Photoionization, $H_D \gamma_D \to X  C$, is the inverse process of radiative recombination and heats the dark plasma. We only consider ionization from the ground state of dark hydrogen. We also ignore photo-excitation of hydrogen because the atom would spontaneously decay back to the ground state, resulting in no net change. 
The cross section of photoionization is \cite{rybicki_lightman_1979}
\begin{eqnarray}
\label{eq:sigmaPhoto}
\sigma_{\rm photo}(\omega) &=& \frac{2^5 \pi^2 \alpha_D^7 m_C^2}{3\omega^4} \frac{e^{-4(\arctan \tau)/\tau}}{1 - e^{-2\pi/\tau}} \nonumber \\
&=&3 \times 10^{-18} \, {\rm cm}^2 \left(\frac{\alpha_D}{10^{-2}}\right)^7 \left(\frac{m_C}{511\,{\rm keV}}\right)^2 \left(\frac{30\,\rm eV}{\omega}\right)^4\left(\left(\frac{e^{-4(\arctan \tau)/\tau}}{1 - e^{-2\pi/\tau}}\right)/0.02\right),
\end{eqnarray}
where $\tau = \left(\omega/\omega_0 -1\right)^{1/2}$ and $\omega_0$ is the dark hydrogen ionization energy in units of eV, which is equal to the Rydberg energy $\frac{1}{2} \alpha_D^2 m_C$. 
The process happens at a rate
\begin{equation}
R_{\rm photo} = n_{H_D} \int_{\omega_0}^\infty 4\pi \sigma_{\rm photo}(\omega) \frac{i(\omega)}{\omega}  d\omega,
\label{eq:phorate}
\end{equation}
where $i(\omega)$ is the intensity of the dark CMB.
The rate of energy transfer is given by
\begin{equation}
\label{eq:phoenrate}
P_{\rm photo} = -n_{H_D} \int_{\omega_0}^\infty 4\pi \sigma_{\rm photo}(\omega) \frac{i(\omega)}{\omega}   \left[  \omega - \frac{1}{2} \alpha_D^2m_C\right] d\omega,
\end{equation}
where the minus sign denotes that this process adds energy to the dark plasma.

\section{Validity of Results}
\label{sec:approx}

In this section we discuss each of the assumptions that we have made in the previous two sections and determine the region of the parameter space in which these assumptions are valid. Throughout our derivations, we assume $\alpha_D \ll 1$ so that perturbative calculations are valid.

\subsection{Ionization and Non-relativistic Electrons}

When dark atoms (formed during recombination before galaxy formation) initially fall into the overdense region in the early Universe, they are shock-heated to a high virial temperature, which is estimated to be~\cite{Fan:2013yva}
\beq
T_{\rm vir} = \frac{G_N M \mu}{5 R_{\rm vir}} \approx 5 \times 10^5~{\rm K} \frac{M}{M^{\rm gal}_{\rm DM}} \frac{m_X}{1~{\rm GeV}} \frac{110~{\rm kpc}}{R_{\rm vir}}.
\label{eq:Tvirial}
\eeq
where $M$ stands for the mass of the virial cluster and $M_{\rm DM}^{\rm gal} = 10^{12} M_\odot$ is the fiducial value for the mass of dark matter in the Milky Way galaxy. $\mu = \rho/n$ is the average mass of a particle in the dark plasma. Assuming that about equal numbers of $X$ and $C$ are present in early galaxies, $\mu = (m_X + m_C)/2 \approx m_X/2$ provided $m_C \ll m_X$. We assume that the virial temperature is high enough so that the dark atoms are ionized entirely and we only need to consider ion-ion and ion-atom scattering processes. This amounts to the  requirement $T_{\rm vir} >$ Ry, which can be translated into an upper bound on $m_C/m_X$:
\beq
\frac{m_C}{m_X} < 8.6 \times 10^{-4} \frac{M}{M^{\rm gal}_{\rm DM}} \frac{110 \mbox{ kpc}}{R_{\rm vir}} \left(\frac{10^{-2}}{\alpha_D}\right)^2. 
\label{eq:upper}
\eeq 
Throughout the paper, we assume $m_C \ll m_X$ and approximate the reduced mass of the atom by $m_C$. 
Note that if dark atoms were not ionized, they could still cool through purely atomic processes, which will be saved for future work. 

The derivations of the cooling functions presented in Section 2 also assume that the dark electrons are non-relativistic. This is true when the temperature of the virialized electrons is below $$T_{\rm rel} \approx 3\times 10^9 \mbox{K} \left(\frac{m_C}{511 \mbox{KeV}}\right).$$
If $T_{\rm vir} < T_{\rm rel}$, it is valid to treat the dark electrons as non-relativistic particles in all the subsequent cooling processes. 
This leads to a lower bound on the mass ratio $m_C/m_X$: 
\begin{equation}
10^{-7} \frac{M}{M^{\rm gal}_{\rm DM}} \frac{110 \mbox{ kpc}}{R_{\rm vir}} < \frac{m_C}{m_X}.
\label{eq:lower}
\end{equation}

\subsection{Cooling in Equlibrium}
Before dark recombination, it is mostly the light ions, the $C$ particles, that lose energy through bremsstrahlung and Compton scattering. The heavy ions $X$ could cool through Rutherford scattering on the $C$ particles. 
The cross section of Rutherford scattering between different ions in the dark plasma is given by 
\beq
\sigma_{R}&=& \frac{8\pi \alpha_D^2}{m_{C}^2v^4}\ln\left(1+\frac{m_{C}v_{C}^2}{\alpha_D}b_{\rm max}\right),  \\
&\approx& \frac{8\pi \alpha_D^2}{(3T)^2}\ln\left(1+\frac{3T}{\alpha_D}b_{\rm max}\right),
\label{eq:rutherford}
\eeq
where, to get the second line, we used the fact that the kinetic energy is set by the temperature of the plasma, $m_C v^2 \approx 3 T$. Notice that the cross section (Eq.~\ref{eq:rutherford}) is universal for $CC$, $XX$ and $XC$ scatterings as long as all ions are in thermal equilibrium and share a common temperature $T$. $b_{\rm max}$ is the maximal impact parameter leading to an effective scattering. Very roughly, we will take $b_{\rm max} = 1/n_C^{1/3}$. For simplicity, we take $n_C = n_X$ given by
\beq
n_C = n_X \approx 7\times 10^{-5}\,{\rm cm}^{-3}\left( \frac{\epsilon}{0.01}\right)\left( \frac{1 \, {\rm GeV}}{m_X}\right)\frac{M}{M^{\rm gal}_{\rm DM}}\left(\frac{110~{\rm kpc}}{R_{\rm vir}}\right)^3,
\label{eq:number}
\eeq
where $\epsilon = 0.01$ is the fraction of energy density in dissipative dark matter, compared to the total dark matter. In deriving this, we assumed that dissipative dark matter spreads uniformly in a 110 kpc radius virial cluster.

The rate of energy equilibration through Rutherford scattering is~\cite{Spitzer1941}
\beq
\frac{P_{\rm eq}}{n_Xn_C} = \frac{2 \sqrt{\pi} \alpha_D^2}{m_X}  \left(\frac{m_C}{T}\right)^{1/2} \ln\left(1+\frac{3T}{\alpha_D n_C^{1/3}}\right). 
\eeq
In deriving the formula, we assume that the energies of electron and proton are always of the same order. 
If the rate is larger than the bremsstrahlung rate in Eq.~\ref{eq:brem}, the cooling of the heavy particles happen adiabatically.  
At $T_{\rm vir}$ in Eq.~\ref{eq:Tvirial}, this turns into a lower bound on the mass ratio $m_C/m_X$:
\beq
\frac{m_C}{m_X} \gtrsim 2 \times 10^{-5}\sqrt{\frac{\alpha_D}{10^{-2}}\frac{M}{M^{\rm gal}_{\rm DM}}\frac{110~{\rm kpc}}{R_{\rm vir}}},
\label{eq:lower2}
\eeq
where we approximate the log factor as 1.

\subsection{Dark Plasma as Fluid}

The easiest way to add the dissipative dark sector to a hydrodynamical simulation is to include it as an additional fluid component. In this section, we identify parameter space in which the dark plasma can be treated as a fluid. We require the mean free path of charged particles to be smaller than the resolution of the state-of-art hydrodynamical simulation.
The mean free path for Rutherford scattering of charged particles in plasma is 
\beq \label{eq:meanfree}
{\ell}&=&\frac{1}{\sigma_R n_C} = \frac{9T^2}{8\pi \alpha_D^2 n_C} \left(\ln\left(1+\frac{3T}{\alpha_D}b_{\rm max} \right)\right)^{-1}, \\
&\approx& 10^{-3} \, {\rm pc}\,\left( \frac{{\rm cm}^{-3}}{n_C}\right) \left(\frac{T}{10^6\, {\rm K}}\right)^2 \left(\frac{10^{-2}}{\alpha_D} \right)^2\left(\frac{21}{\ln\left(1+\frac{3T}{\alpha_Dn_C^{1/3}} \right)}\right), \\
&\approx& 3.7 \, {\rm pc}\,\left( \frac{0.01}{\epsilon}\right) \left(\frac{m_X}{1 \, {\rm GeV}}\right)^3\left(\frac{10^{-2}}{\alpha_D} \right)^2\frac{M}{M^{\rm gal}_{\rm DM}}\,\frac{R_{\rm vir}}{110~{\rm kpc}},
\eeq
where in the last line, we used the number density in Eq.~\ref{eq:number} and the initial virial temperature in Eq.~\ref{eq:Tvirial}. We assumed that dissipative dark matter spreads uniformly in a 110 kpc radius virial cluster. The resolution of current simulations is below 100 pc. For example, the smallest scale over which the hydrodynamics is resolved is 48 pc in Illustris simulation~\cite{Vogelsberger:2014kha, Vogelsberger:2014dza}. Requiring $\ell < 50$ pc, we find that the fluid approximation is valid if 
\beq
\left( \frac{0.01}{\epsilon}\right) \left(\frac{m_X}{1 \, {\rm GeV}}\right)^3\left(\frac{10^{-2}}{\alpha_D} \right)^2\frac{M}{M^{\rm gal}_{\rm DM}}\,\frac{R_{\rm vir}}{110~{\rm kpc}} < 13.5.
\label{eq:fluid}
\eeq
For $X$ with mass above GeV and $\alpha_D \ll 10^{-2}$, the fluid approximation breaks down and calls for new ways to include them in a simulation, which we will not explore here.


There are a few comments in order:
\begin{itemize}
{\item Our estimate is conservative. If the dissipative dark matter spatial distribution were concentrated in a smaller region (with radius less than 110 kpc), the number density would be larger and the mean free path would be even shorter. }
{\item We estimate the mean free path for a virialized halo. What about particles in the intergalactic medium that have not fallen into an overdense region? Today the critical density is $\rho_c \approx 5 \times 10^{-6}~{\rm GeV}/{\rm cm}^3$. So in the intergalactic medium the number densities are $n \sim \epsilon \rho_c/m_X \sim 5 \times 10^{-8}~{\rm cm}^{-3}$. On the other hand, the temperatures are also much lower. If we plug in the dark CMB temperature of order 1 K, we see that they have $\ell$ even smaller than inside the halo. Their density is lower but the temperature is much lower still, and this increases the scattering rate.}
{\item Nonrelativistic particles have a temperature that changes with $z$ in a different manner than a relativistic gas. 
The temperature of nonrelativistic gas changes as $T \sim p/\rho \sim 1/a^2$. Thus, if at early times the $X$ and $C$ particles were in equilibrium with the CMB, they will be colder than the CMB now and our estimate that they have small mean free path is safe since $\ell \propto T^2$.}
{\item Assuming the simulations start from $z \sim 100$, we also want to check if the fluid approximation is valid for early times. Also, we should check whether it is valid at late times for particles that do not have enhanced density from falling into a halo. At early times, Compton scattering is also important. The mean free path for Compton scattering is 
\beq
\ell &=& \frac{1}{\sigma_T n_\gamma} = \frac{3 m_C^2}{8\pi\alpha_D^2 n_\gamma} \\
&\approx & 0.005 \, {\rm pc} \left(\frac{m_C}{511 \, {\rm KeV}}\right)^2 \left(\frac{10^{-2}}{\alpha_D}\right)^2 \left(\frac{101}{(1+z)}\right)^3,
\eeq
where in the second line we used $n_\gamma=2\zeta(3)(T_{D}^0 (1+z))^3/\pi^2 $ with the current dark CMB temperature, $T_D^0$, half of our CMB temperature 2.7 $K$. So at early times the electrons ($C$ particles) are Compton scattering frequently off the CMB; this holds until $z \sim 10$. For the heavy field $X$, Rutherford scattering still dominates.}
{\item If we extrapolate back in time, $n \propto (1+z)^3$, whereas $T \propto (1+z)^2$ while the particles are kinetically decoupled from the CMB (behaving as nonrelativistic particles in an adiabatically expanding universe) or $T \propto (1+z)$ if the particles are interacting frequently with the CMB. Being conservative, if we put in $T \propto (1+z)^2$ beginning at $T = 1$ K now, and $n$ growing relative to $\epsilon$ times the critical density, the estimate shows that $\ell$ is still small at $z = 100$.}
{\item When a considerable fraction of ions are recombined into dark hydrogen atoms, dark electrons could also scatter off dark atoms elastically with a cross section
\beq
\sigma_{{\rm elastic}} &=& \frac{7 \pi}{3 m_C^2 v^2} \times z\left(\frac{v^2}{\alpha_D^2}\right), \nonumber \\
&\approx& \frac{7 \pi}{9 m_C T} \times z\left(\frac{3T}{\alpha_D^2 m_C}\right), \\
{\rm with} \quad z(x) &=& \frac{x \left(x^2+18x/7+12/7\right)}{(1+x)^3}. \nonumber
\eeq
This was obtained by leading order Born approximation. When $T \sim \alpha_D^2 m_C/2$, dark atoms start to form and the function $z$ gives an order one dimensionless number. The mean free path for the $C-H_D$ elastic scattering is then
\beq
\ell &=& \frac{1}{\sigma_{{\rm elastic}} n_{H_D}} = \frac{9 m_CT}{7 \pi n_{H_D}} \frac{1}{z\left(\frac{3T}{\alpha_D^2 m_C}\right)}  \nonumber \\
&\approx &44\, {\rm pc}\, \left(\frac{x}{z(x)}\right)\,\left(\frac{\alpha_D}{10^{-2}}\right)^2 \left(\frac{m_C}{511\,{\rm keV}}\right)^2 \left(\frac{0.01}{\epsilon}\right)\left(\frac{m_X}{1 \, {\rm GeV}}\right) \frac{M^{\rm gal}_{\rm DM}}{M}\left(\frac{R_{\rm vir}}{110~{\rm kpc}}\right)^3\frac{1}{r},
\eeq
where $x = 3T/(\alpha_D^2 m_C)$ and $r$ is the fraction of $C$ particles that are recombined into atoms and in the second line, we used Eq.~\ref{eq:number}. Thus when the fraction of ions is order one, Rutherford scattering still dominates. When most ions are recombined into atoms, the mean free path of the electrons (scattering with the atoms) is below 1 pc for $\epsilon = 0.01, \alpha_D = 0.01$ and $m_X \lesssim 0.036$ GeV. Yet at that point, cooling through ion-atom collisions ceases to be efficient. Further cooling is possible with atomic and molecular processes, which goes beyond the scope of this paper. }
\end{itemize}

\subsection{Dark Plasma is Optically Thin}
\label{sec:opticallythin}

In all the processes in which dark photons are emitted, they must escape from the galaxy and carry away energy without being re-absorbed. A dark photon scatters with light $C$ particles, so the mean free path of $\gamma_D$ can be approximated as~\cite{Fan:2013yva}
\beq
\ell= \frac{1}{\sigma_T n_C} = \frac{3 m_C^2}{8\pi \alpha_D^2 n_C} \approx 4 \times 10^6~{\rm kpc} \left(\frac{m_C}{511 \,{\rm keV}}\right)^2 \left(\frac{10^{-2}}{\alpha_D}\right)^2\,\left( \frac{0.01}{\epsilon}\right) \left(\frac{m_X}{1 \, {\rm GeV}}\right)\left(\frac{M^{\rm gal}_{\rm DM}}{M}\right)\left(\frac{R_{\rm vir}}{110~{\rm kpc}}\right)^3,
\eeq
in which we have used the Thomson cross section for $\gamma_D$--$C$ scattering with the number density in Eq.~\ref{eq:number}.
The long mean free path demonstrates that dark photons definitely escape the galaxy at early times. 

When dark recombination happens, one also needs to check how fast the emitted energetic photons are absorbed. Given the photonionization cross section in Eq.~\ref{eq:sigmaPhoto}, the mean free path for photons with energy $\sim$ Ry is 
\beq
\ell &=& \frac{1}{\sigma_{\rm photo} n_{H_D}} \nonumber \\
&\approx& 1.6 \,{\rm kpc} \left(\frac{10^{-2}}{\alpha_D}\right)^7 \left(\frac{511\,{\rm keV}}{m_C}\right)^2 \left(\frac{\omega}{30\,\rm eV}\right)^4 \left(\frac{0.01}{\epsilon}\right) \left(\frac{m_X}{1\,\rm GeV}\right)\left(\frac{M^{\rm gal}_{\rm DM}}{M}\right)\left(\frac{R_{\rm vir}}{110~{\rm kpc}}\right)^3 \frac{1}{r}\,
\eeq
where the reference number for $\omega$, 30 eV, is chosen to be close to the Rydberg energy for $\alpha_D = 10^{-2}$ and $m_C = 511$ keV. $r$ is the fraction of $C$ particles that are recombined into atoms. When the fraction of ions converting into atoms is large enough and the mean free path is smaller than the galaxy size, the assumption that gas is optically thin breaks down. When that happens, cooling is no longer efficient. 


\subsection{Gaunt Factors} 
Gaunt factors for free-free (bremsstrahlung) and free-bound (recombination) processes have been computed by Ref.~\cite{karzas_latter_1961}. The free-free Gaunt factors have been presented as a function of $\omega/T$, the emitted photon energy divided by the temperature (or equivalently the initial kinetic energy of the dark electrons). They are between 1 and 3 for $10^{-3} < \omega/T < 1$. For even smaller $\omega/T$, the emitted photon contributes negligibly to the kinetic energy loss. Thus, for simplicity, we take the free-free Gaunt factors to be one.

The free-bound Gaunt factor has been computed as a function of $K/{\rm Ry}$ for various energy levels of the dark atom. As depicted in Fig. 19 of Ref.~\cite{karzas_latter_1961}, for all energy levels, it is approximately one when the initial kinetic energy is at or below 2.7 times the Rydberg energy of the dark atom. Simple analytic formulas exist for low energy levels. For instance, for $n=1$, 
\begin{equation}
\label{eq:bfGaunt}
g_{{\rm fb},(n=1)} =  8 \sqrt{3}\pi  \frac{\alpha_D^2}{v^2 + \alpha_D^2} \frac{e^{-4  \arctan \tau/\tau}}{1- e^{-2\pi /\tau}}, \end{equation}
where 
$\tau = \sqrt{\frac{K}{\rm Ry}}.$
From these analytic formulas, we see that free-bound Gaunt factors are about 1 when $K$ is near the Rydberg energy and drop to 0.5 when $K/{\rm Ry} \sim 100$. For even higher temperatures and kinetic energies of the incident electrons, the recombination rate is small anyway and contributes negligibly to the cooling. Thus, it is reasonable to approximate $g_{\rm fb}$ as one.

\subsection{Binary Encounter Approximation}
We used the binary encounter approximation in obtaining rates of collisional ionization.
As discussed in Ref.~\cite{peterkop_1977}, the binary encounter approximation and its variants perform as well as quantum mechanical approximations. The underlying reason for this agreement is that the Rutherford formula, on which the approximation is based, holds in both classical and quantum mechanics. 

In the binary encounter approximation, we assume that the effect of the nucleus is negligible, or that its effect is simply to accelerate the free electron (in Burgess' improvement). Essentially, we assume that the electric force between the free incident electron and the bound electron is much greater than that between the nucleus and the bound electron. This assumption amounts to $b \lesssim a_0,$ where $b$ is the impact parameter. This is a self-consistent assumption and also agrees with the physical intuition that collisions at large impact parameter will not lead to ionization. One can check that the maximal impact parameter leading to collisional ionization is about $a_0$, performing the calculation within the classical binary-encounter approximation. The resulting cross section in Eq.~\ref{eq:BE} is also consistent with the approximation. Using the classical binary approximation, the cross section takes the maximal value of $\pi a_0^2$ when $x\equiv \alpha_D^2/v^2=1/2$. The effective impact parameter $b_{\rm eff}$, defined by $\sigma_{\rm ion, binary} = \pi b_{\rm eff}^2$, is thus always smaller than or at most equal to $a_0$. 

In principle, one could solve the Schr${\ddot{\rm o}}$dinger equation numerically to get a precise result on a more solid ground. Yet the collisional breakup of a bound state of two particles in a system of three charged particles turns out to be a very difficult problem. The key issue is the wave function for systems with three or more charged particles in the presence of the long-range Coloumb force was unknown. The problem has been tackled successfully numerically by Rescigno et al. in Ref.~\cite{Rescigno1999} and by Barlett in Ref.~\cite{barlett} using the so-called ``exterior complex scaling'' method. Yet the method is computationally intensive and one has to solve systems of complex linear equation on the order of 5 million by 5 million \cite{Rescigno1999}. Thus we do not pursue this direction.



\subsection{Born Approximation and BE Scaling}

From the discussion in Section~\ref{sec:collisionalExcitation}, one can see that in general, the first-order Born cross section for different collisional atomic transitions can be written systematically as 
 \beq
 \sigma_{\rm Born} = 4 \pi a_0^2 \,\frac{\rm Ry}{K} \, F_{\rm Born}\left(\frac{\rm Ry}{K}\right), 
 \eeq
where $F_{\rm Born}$ is the collision strength (which needs to be multiplied by a constant to be consistent with the standard definition of the collision strength).  
Strictly speaking, the Born approximation is valid at large incident energy $K \gg$ Ry. It over-estimates the cross section by an order one number at $K \sim$ Ry.  This is corrected by BE scaling in Eq.~\ref{eq:BEscaling}. At large $K$, BE scaling doesn't change the result from the Born approximation significantly. Yet it reduces the cross section at lower $K$ and shifts the peak of the cross section to a higher $K$. The qualitative justification, similar to Burgess' improvement for the collisional ionization, is that the ``effective'' incident energy seen by the bound electron is $K$ plus the potential energy of the bound electron. So far BE scaling is not derived from first principles. Thus the combination Ry$+\Delta E$ in the BE scaling equation cannot be taken as a rigid rule. It only serves as an indicator of the order of magnitude of a constant shift to be added to the kinetic energy of the incident electron.

\subsection{Neglecting Proton-hydrogen Collisions}
\label{sec:neglectProtons}
The dark proton-impact collisional processes is most effective when the incident proton's velocity is on the order of $\alpha_D$, giving a cross section of order $\pi a_0^2$. Yet if the protons are in thermal equilibrium with the electrons, their average velocity, $v \sim \sqrt{\frac{3T}{m_X}}$, is much less than $\alpha_D$ at low temperatures when there are hydrogen atoms present. At velocities much smaller than $\alpha_D$, the inelastic collisional cross sections are several orders of magnitude below $\pi a_0^2$ and are thus negligible compared to those of electron-impact collisional processes~\cite{Bates1953,Vriens1966, Tseliakhovich:2012hb}. 
This can be understood heuristically in the classical picture. Consider the case of collisional ionization. Imagine a dark electron in a stable orbit around a dark proton. A free proton approaches the electron slowly. As it approaches, the electron falls into the potential well of the new proton. Because the proton approaches slowly, the relative velocity between the electron and the incident proton is not great enough for the electron to escape the potential well. The electron becomes bound to the incident proton and is dragged along with it. This happens as long as the proton is moving at a speed smaller than the escape velocity of the electron, which is $\alpha_D$. Thus, as long as $T \ll m_X \alpha_D^2,$ we can ignore proton-initiated impacts. At higher temperatures, $T \gtrsim m_X \alpha_D^2$, the proton-impact collisional ionization process is turned on. Yet all of the hydrogen atoms have already been ionized at such high temperatures. Thus $n_{H_D} \approx 0$ and it is safe to ignore the proton-impact processes.

\subsection{Neglecting Other Collisions}

We have also neglected $C-C$ and $X-X$ collisions. The reasons for neglecting these processes are the same as for neglecting them in discussing cooling of ordinary baryons. For example, consider electron-electron collisions. As summarized by Ref. \cite{haug_1975}, in the non-relativistic limit and in the rest frame of one of the electrons, the cross-section for emitting a photon of energy $\omega$ due to bremsstrahlung off of another electron is
\begin{equation}
\frac{d\sigma}{d\omega} = \frac{4}{15}\frac{\alpha_D a_0^2}{\omega} F\left(\frac{8\omega}{m_Cv^2}\right),
\end{equation}
where
\begin{equation}
F(x) = \left[17 - \frac{3x^2}{(2-x)^2}\right]\sqrt{1-x} + \frac{12(2-x)^4 - 7x^2(2-x)^2 -3x^4}{(2-x)^3} \ln \frac{1 + \sqrt{1-x}}{\sqrt x}.
\end{equation}
The energy loss rate due to electron-electron bremsstrahlung is
\begin{equation}
\begin{split}
\frac{P}{n_C^2} &\sim \int d\omega \left\langle \frac{d\sigma}{d\omega} \omega v \right\rangle \\
&\sim \alpha_D a_0^2 \left(\frac{m_C}{T}\right)^{3/2}  \int_0^\infty dv v^3 e^{-\frac{m_Cv^2}{2T}} \int_0^{\frac{1}{8}m_Cv^2} d\omega F\left(\frac{8\omega}{mv^2}\right)\\
&\sim  a_0 \left(\frac{m_C}{T}\right)^{3/2}  \int_0^\infty dv v^5 e^{-\frac{m_Cv^2}{2T}}\\
&\sim a_0 \left(\frac{T}{m_C}\right)^{3/2},
\end{split}
\end{equation}
which is negligible by our non-relativistic electrons assumption, in which $T \ll m_C$.

\section{Conclusions and Outlook}
\label{sec:conclusion}

Given the little knowledge we have about dark matter, it is important to explore new possible dark matter models. In this paper, we consider a multi-component dark matter scenario with a sub-dominant component dissipating energy, analogous to ordinary baryons. Specifically we consider the simplest possibility, in which the dissipative dark sector consists of a dark electron and proton both charged under a gauged $U(1)$. We have computed cooling functions, including energy dissipation rates and rates of processes that change the number densities of different species, for the following processes: Compton scattering, bremsstrahlung, recombination, collisional ionization and collisional excitation. We also consider photoionization which heats the dark plasma. This paper is the first publication of analytic formulas for the cooling processes, other than Compton scattering and bremsstrahlung, in the dark matter literature, which includes dependence on the dark electron's mass, the fine structure coupling, the temperature, and other relevant parameters. We also discuss the approximations we rely on to derive the formulas and identify the parameter space in which our results are applicable (more specifically, Eq.~\ref{eq:lower},~\ref{eq:lower2}, \ref{eq:upper}, \ref{eq:fluid}). The key results are summarized in Table~\ref{table:rates}. These cooling functions can be fed into numerical simulations including a subdominant dissipative component to get a more definite answer as to whether and how a dark disk could be formed and estimate important properties of a possible dark disk such as its height and surface density. The numerical results could be further compared with the upcoming Gaia data, hopefully to settle the debate on whether a dark disk formed from a dissipative component is allowed. 

In this article, we assume that dark atoms falling into galaxies get ionized initially and cool through ion-ion and ion-atom collisional processes. We do not consider the parameter space with a large binding energy so that the dark atoms are not ionized by shock waves. We also do not consider further cooling through atomic and molecular processes once the dark ions are entirely recombined into atoms.\footnote{The importance of hyperfine transition has been studied in the dark matter model with dark atoms being the only component~\cite{Boddy:2016bbu}.} These processes could be important to study the formation of smaller compact objects such as dark stars. We leave them for future work.

\begin{table}[h!]
\centering
	\begin{tabular}{| c | c | c |}
	\hline
	Processes & Energy loss (gain) rate & Process rate	\\ \hline
	Compton & Eq.~\ref{eq:compton} & N/A	\\ 
	Bremsstrahlung	 & Eq.~\ref{eq:brem} & N/A \\ 
	Recombination & Eq.~\ref{eq:rateofenergylossrec} & Eq.~\ref{eq:recombRate}	\\ 
	Collisional Ionization	 & Eq.~\ref{eq:ionenergyrate} & Eq.~\ref{eq:ionrate} \\ 
	Collisional Excitation & Eq.~\ref{eq:1s2pPower} & N/A	\\ \hline
	Photoionization & Eq.~\ref{eq:phoenrate} & Eq.~\ref{eq:phorate} \\ \hline
	\end{tabular}
\caption{Summary of the cooling functions. Except for photoionization (the last line), all of the processes reduce the dark plasma's energy. We include the rates of processes such as recombination and collisional ionization that change the number densities of ions and dark atoms. } \label{table:rates}
\end{table}

\section*{Acknowledgments}
We thank Seth Ashman, Francis-Yan Cyr-Racine, Ian Dell'Antonio, Ricky Oliver and Matt Reece for useful discussions and comments. We thank Shuyi Li for participating in the initial stages of this project.
ENR gratefully acknowledges support from an UTRA/NASA Space Grant Summer Scholarship, co-funded by Brown University and the NASA RI Space Grant. JF is supported by the DOE grant DE-SC-0010010.

\bibliography{references}
\bibliographystyle{utphys}

\end{document}